\documentstyle[11pt,amscd]{article}                                    
\newcommand{\frak}[1]{{\mathbf #1}}                                  

\def\AFOUR{%
\setlength{\textheight}{9.0in}%
\setlength{\textwidth}{5.75in}%
\setlength{\topmargin}{-0.375in}%
\hoffset -.5in%
\renewcommand{\baselinestretch}{1.17}%
\setlength{\parskip}{6pt plus 2pt}}%
\AFOUR                                          
\def\car{\mathop{\square}}
\def\carre#1#2{\raise 2pt\hbox{$\scriptstyle #1$}\car_{#2}}
\makeatletter
\def\section{\@startsection {section}{1}{\z@}{-3.5ex plus -1ex minus
 -.2ex}{2.3ex plus .2ex}{\large\bf}}
\def\subsection{\@startsection{subsection}{2}{\z@}{-3.25ex plus -1ex 
minus
 -.2ex}{1.5ex plus .2ex}{\normalsize\bf}}
\makeatother
\newcommand{\nc}{\newcommand}
\newcommand{\rnc}{\renewcommand}
\nc{\be}{\begin{equation}}
\nc{\ee}{\end{equation}}
\nc{\bea}{\begin{eqnarray}}
\nc{\eea}{\end{eqnarray}}

\def\href#1#2{{#2}}

\rnc{\a}{\alpha}
\nc{\ab}{\bar{\a}}
\nc{\ap}{\a^{+}}
\nc{\abm}{\ab^{-}}
\rnc{\b}{\beta}
\nc{\bb}{\bar{\b}}
\nc{\bbp}{\bb_{\zb}^{+}}
\nc{\bm}{\b_{z}^{-}}
\nc{\oa}{\overline{\a}}
\nc{\ob}{\overline{\b}}
\rnc{\gg}{\gamma}
\rnc{\d}{\delta}
\nc{\f}{\phi}
\nc{\fb}{\bar{\phi}}
\nc{\vf}{\varphi}
\nc{\p}{\psi}

\rnc{\c}{\chi}
\nc{\la}{\lambda}
\nc{\m}{{\mathrm m}}
\nc{\n}{\nu}
\rnc{\o}{\omega}
\nc{\Om}{\Omega}
\rnc{\t}{\theta}
\nc{\eps}{\epsilon}
\rnc{\S}{\Sigma}
\nc{\F}{\Phi}
\nc{\trac}[2]{{\textstyle\frac{#1}{#2}}}
\nc{\ex}[1]{\mbox{e}^{\,\textstyle#1}}
\nc{\mat}[4]{\left(\begin{array}{cc}#1&#2\\#3&#4\end{array}\right)}
\nc{\som}[9]{\left(\begin{array}{ccc}#1&#2&#3\\#4&#5&#6\\#7&#8&#9%
\end{array}\right)}
\nc{\tr}{\mathop{\mbox{tr}}\nolimits}
\nc{\ad}{\mathop{\mbox{ad}}\nolimits}
\nc{\Tr}{\mathop{\mbox{Tr}}\nolimits}
\nc{\Det}{\mathop{\mbox{Det}}\nolimits}
\nc{\rk}{\mathop{\mbox{rk}}\nolimits}
\nc{\ra}{\rightarrow}
\nc{\Ra}{\Rightarrow}
\nc{\LRa}{\Leftrightarrow}
\nc{\ot}{\otimes}
\rnc{\ss}{\subset}
\nc{\nul}{\noindent\underline}
\nc{\non}{\nonumber\\}
\nc{\subs}[1]{{\vspace*{0.5cm}}%
{\vspace*{0.3cm}}}
\nc{\zb}{\bar{z}}
\rnc{\lg}{\frak{g}}
\nc{\lt}{\frak{t}}
\nc{\lk}{\frak{k}}
\nc{\lh}{\frak{h}}
\nc{\pik}{\Pi_{\lk}}
\nc{\pip}{\Pi_{+}}
\nc{\pim}{\Pi_{-}}
\nc{\pih}{\Pi_{\lh}}
\nc{\jz}{J_{z}}
\nc{\jzh}{\jz^{\lh}}
\nc{\jzp}{\jz^{+}}
\nc{\jzm}{\jz^{-}}
\nc{\del}{\partial}
\nc{\dz}{\del_{z}}
\nc{\dzb}{\del_{\bar{z}}}
\nc{\az}{A_{z}}
\nc{\azb}{A_{\bar{z}}}
\nc{\g}{g^{-1}}
\nc{\dw}{\Delta_{W}}
\nc{\Ad}{{\mbox{Ad}}}
\nc{\ks}{Ka\-za\-ma-\-Su\-zu\-ki}
\nc{\KS}{\ks}
\nc{\ksm}{\ks\ model}
\rnc{\AA}{{\Bbb A}}
\nc{\BB}{{\Bbb B}}
\nc{\CC}{{\Bbb C}}
\nc{\PP}{{\Bbb P}}
\nc{\cpm}{\CC\PP(m)}
\nc{\cpn}{\CC\PP(n)}
\nc{\cp}[1]{\CC\PP(#1)}
\nc{\gmn}{G(m,m+n)}
\nc{\gmnk}{\gmn_{k}}
\nc{\cO}{{\cal O}}
\nc{\bcO}{\bar{\cO}}
\nc{\bO}{\bar{O}}
\nc{\oQ}{\overline{Q}}
\nc{\ie}{{\it i.e.~}}
\nc{\eg}{{\it e.g.~}}
\begin{document}
\global\parskip 4pt
\makeatother\begin{titlepage}
\begin{flushright}
{ROM2F/99-47}
\end{flushright}
\vspace*{0.5in}
\begin{center}
{\LARGE\sc  Type I Vacua from Diagonal $Z_3$-Orbifolds}
\vskip 0.8cm
\makeatletter

\begin{tabular}{c}
{\bf\large Gianfranco Pradisi\footnotemark }
\\
\\
\\
Dipartimento di Fisica \\
Universit\`a di Roma ``Tor Vergata''\\
and \\ 
INFN, sezione di Roma ``Tor Vergata'' \\
Via della Ricerca Scientifica, 1 - 00173 Rome, ITALY\\
\end{tabular}

\end{center}
\addtocounter{footnote}{0}%
\footnotetext{e-mail: gianfranco.pradisi@roma2.infn.it}
\vskip .50in
\begin{abstract}
\noindent
We discuss the open descendants of diagonal irrational $Z_3$ orbifolds,
starting from the $c=2$ case and analyzing six-dimensional and
four-dimensional models.  As recently argued, their consistency is linked to
the presence of geometric discrete moduli.  The different classes of open
descendants, related to different resolutions of the fixed-point ambiguities,
are distinguished by the number of geometric fixed points surviving the
unoriented projection.
\end{abstract} 
\makeatother 
\end{titlepage}

\setcounter{footnote}{0}

\section{Introduction}

Open superstrings models descend from oriented closed 
superstrings \cite{cargese,ps,bs,toroidal}, 
and display the same level of consistency of their heterotic counterparts. 
However, for several years they have been considered less attractive, 
since it was believed that 
weakly-coupled heterotic superstrings compactified on Calabi-Yau threefolds
could provide a natural bridge between a unified theory of all interactions 
and the Standard Model \cite{dien}. Even the unusual low-energy properties
of open-string models \cite{bs}, a first clear indication
of their potential, were not immediately 
appreciated. This viewpoint has dramatically changed since 
weak/strong coupling dualities have effective led to the unification
of all five 
known superstrings (heterotic, Type I and Type II), that together with 
the eleven-dimensional supergravity are now regarded as 
different asymptotic expansions of 
M-theory \cite{mtheo}.  Moreover, the 
perturbative oscillations of 
D-branes, BPS solitons carrying Ramond-Ramond 
charges \cite{pol} that play a central role in the web of string dualities, are 
described by open superstrings.  These findings have 
completely changed our picture of 
how the Standard Model should be embedded within string theory or M-theory. 
To wit, while in a heterotic susy-GUT scenario the string scale and the
Planck scale are essentially identified, in Type I 
vacua both the string scale and the dimensions felt by gauge interactions 
could be as low as a few TeV. Moreover, in Type I
vacua the gauge fields generally invade only some of the dimensions, while
the remaining ones, felt only via gravitational interactions, could
be far larger, and even of millimeter size \cite{tevscale}.  
In this new Kaluza-Klein scenario, often called ``Brane World'' 
\cite{kakutye}, several 
fundamental issues like the gauge hierarchy problem and the nature of
supersymmetry breaking have to be 
reconsidered. At any rate, Type I vacua have a central role in this 
setting, but only a limited number of them has been explored so far.

In this paper we construct a class of six and four dimensional 
Type I vacua from ``non-geometric'' $Z_3$ orbifolds of the parent Type IIB 
theory.  The more familiar ``geometric'' Type I vacua are open descendants 
of Type IIB models obtained modding out the closed spectrum with the  
world-sheet parity operator $\Omega$, that interchanges left and right 
moving sectors, and adding suitable twisted states, unoriented open 
superstrings.  From a Conformal Field Theory point of view, their 
construction translates into a set of rules based on 
{\it sewing constraints} that, for a given left-right symmetric model, 
allow in general several possible 
descendants \cite{sonoda,lew,fps,pss1,pss2}.  
One has indeed the freedom of changing the unoriented truncation of the closed 
sector (the Klein-bottle amplitude) compatibly with the 
{\it ``crosscap constraint''} \cite{fps,pss1,pss2,susy95} and of adding suitable
boundary-states, that  may or may not \cite{fusc}
respect all the symmetries of the 
bulk.  The final ingredient is the  solution of tadpole conditions
that fix (partly) the Chan-Paton gauge  groups.  
This brings about a number of surprises.
For instance, one can have Type I models without open strings 
\cite{gepner,dapa}, while unconventional 
Klein-bottle projections can even require that supersymmetry be
broken to lowest order in the open sector 
\cite{ads,aadds} as a result of the simultaneous presence of branes and
anti-branes.   These last models are the 
first in which supersymmetry breaking is a consistency 
condition rather than an option.  

There is another possibility, that consists in ``dressing'' $\Omega$ with the 
action of involutions $\cal{I}$ of order two.  Again, from a Conformal 
Field Theory point of view, this is equivalent to applying 
the construction not to the ``geometric'' Type IIB, but to a 
different
parent theory based on a different, non geometric, GSO projection. Rational
models of this type were studied very early in \cite{bs}, but a closer
look at the relation between the different approaches can be very useful,
and in particular shed some 
light on the generalization of the intuitive concepts 
of D-branes and O-planes to non geometric settings \cite{fusc,resho}.      

To be less generic, let us 
consider the open descendants of the  Type IIB
superstring compactified on a $Z_3$ orbifold to $D=6$.  The ``geometric''
torus partition function corresponds to the so-called
``charge conjugation'' modular  invariant.  The open descendants of this model
have been constructed by several authors both at rational {\cite{gepner}} and at
generic points of moduli space \cite{dapa,gijo}.  Recently, another class 
of open descendants was constructed in \cite{bgk,ab} 
combining  $\Omega$ with a conjugation of the complex internal 
coordinates (and with a corresponding action on the fermions).  The 
resulting models, already known as open descendants of Gepner
models \cite{gepner}, {\it i.e.} at special points of moduli space, 
exhibit some interesting properties.  
First, the spectrum includes twisted open strings, and this feature, 
common to other non-geometric Type I vacua \cite{noi}, can
be interpreted in terms 
of D$7$-branes at angles.  Second, their 
consistency rests on the presence of quantized ``geometric'' moduli (the
off-diagonal components of the target-space metric). These are also 
responsible for the rank reduction of the Chan-Paton groups \cite{ab}, much in
the same way as the $B$-field (now a continuous deformation) is in the
conventional case \cite{toroidal,zurB,carlo}.  In this paper we explore
further the world sheet structure of these models, and show how they 
may be regarded as open descendants of 
theories whose GSO projections correspond to 
``diagonal'' modular invariants. We also extend the 
construction to four dimensional models that have similar properties.   

The plan of the paper is the following. 
In Section 2 we begin by discussing the open descendants of the bosonic
$c=2$ diagonal 
$Z_3$-orbifold.  This is useful to address the behavior of the 
irrational deformations.  In Section 3 we recover 
the six-dimensional models of 
\cite{bgk,ab} and extend the procedure to investigate additional models in 
four dimensions.  Section 4 is devoted to our conclusions 
and to some conjectures.  Notation and conventions are
illustrated in Appendix A.

\section{The $c=2$ diagonal $Z_3$-orbifold and its open descendants}

A pair of bosonic fields compactified on a $Z_3$-orbifold of a two-torus can be
described by a single complex field $Z$ modded out by the action of the orbifold
group that, in the $k$-th twisted sector, is
\be
Z(\sigma + 2\pi, \tau) \ = \ \omega^k \ Z(\sigma, \tau) \ \qquad,
\label{orbgroup}
\ee
where $\omega = e^{2 i \pi/3}$.  The partition function of the resulting
theory can be written in terms of a sesquilinear combination of chiral blocks
(see Appendix A).  In particular, the geometric orbifold corresponds to
the so called ``charge conjugation'' modular invariant
\bea
{\cal T}_{\it{cc}} \ = \ {1 \over 3} &\!\!\!\!{\bigg[}&\!\!\!\! \Phi_{00}\bar\Phi_{00} \,
\Lambda + \Phi_{01}\bar\Phi_{02} + \Phi_{02}\bar\Phi_{01} \ + \ 3 \
(\Phi_{10}\bar\Phi_{20} + \Phi_{11}\bar\Phi_{22} + 
\Phi_{12}\bar\Phi_{21} \ ) \nonumber \\   &\!\!\!\!+&\!\!\!\! 3 \ 
(\Phi_{20}\bar\Phi_{10} + \Phi_{21}\bar\Phi_{12} + 
\Phi_{22}\bar\Phi_{11} \ ) \ {\bigg]} \qquad, 
\label{cminvc2}
\eea
where $\Lambda$ is the Narain lattice sum and the factors of three are connected
to the three points fixed under the $Z_3$ action.  Denoting by $m_i$ and $n_i$
respectively momenta and winding modes, and letting  
\bea
\bf{p} \ &=& \ {1 \over \sqrt{2 X_2 Y_2}} \ [ \ X m_1 - m_2 - \bar{Y} \ 
( \ n_1 + X  n_2 \ ) \ ] \qquad, 
\nonumber \\
\bf{\tilde{p}} \ &=& \ {1 \over {\sqrt{2 X_2 Y_2}}} \ [ \ X m_1 - m_2 - Y \ 
( \ n_1 + X  n_2 \ ) \ ]
\qquad ,
\label{momenta}
\eea
where $X=X_1 + i X_2$ and $Y=Y_1 + i Y_2$ are the two complex moduli of 
the target two-torus, 
connected with metric and antisymmetric tensor by the relations
\be  
g = {\alpha ' Y_2 \over X_2} \left( \matrix{1 & X_1 \cr X_1 & |X|^2\cr} 
\right) \ , \qquad   B = \alpha ' Y_1 \left( \matrix{0 & 1 \cr -1 & 0\cr}
\right) \ ,
\label{z3metric}
\ee
the lattice sum is
\be
\Lambda \ = \ \sum_{({\bf p}, {\bf {\tilde p}})\in 
\Gamma_{2,2}} \, { q^{\frac{|{\bf{p}}|^2}{2}} \
{\bar{q}}^{\frac{{|\tilde{\bf{p}}|}^2}{2}}} \qquad .
\ee
We set for simplicity $B=0$ ({\it i.e.} $Y_1=0$ ), and once the lattice 
has been normalized to
have basis vectors of length $2$ in $R^2$ units, it is easy to recognize
that a possible choice for a sensible $Z_3$ action is $X_1=-1/2$,
$X_2=\sqrt{3}/2$ and $Y_2=\sqrt{3} R^2 / \alpha '$.  The open descendants of
this geometric orbifold  have been analyzed in several contexts, and result in a
class of theories with only Neumann strings  ({\it i.e.} excitations of 
D25-branes if one refers to
the bosonic critical string theory, or of D9-branes in Type IIB) in the open
sector and with (bulk) twisted sectors flowing in the tree channel, 
including ``massless'' ones.  

In order to produce
a diagonal GSO projection, one has to combine the orbifold action
(\ref{orbgroup})  with an involution that conjugates the eigenvalues of the
right-moving coordinates.  As observed in \cite{bgk}, this diagonal action is
natural on the complex field obtained T-dualizing one of the real
components of the field $Z$, rather than on the field $Z$ itself.  The resulting 
open descendants, when dressed with Type II critical superstring coordinates,
are sometimes referred to as Type I' models, and may be regarded as orientifolds
of the Type IIA superstring \cite{hora}.  Coming back to the bosonic 
model, the torus partition function displays neatly the
diagonal combination  
\bea  {\cal T_{\it{d}}}\ = \
{1 \over 3} &\!\!\!\!{\bigg[}&\!\!\!\! \Phi_{00}\bar\Phi_{00} \, \Lambda +
\Phi_{01}\bar\Phi_{01} + \Phi_{02}\bar\Phi_{02} \ + \ 3 \ ( \
\Phi_{10}\bar\Phi_{10} + \Phi_{11}\bar\Phi_{11} + 
\Phi_{12}\bar\Phi_{12} \ )  \nonumber \\   &\!\!\!\!+&\!\!\!\!  3 \ ( \
\Phi_{20}\bar\Phi_{20} + \Phi_{21}\bar\Phi_{21} + 
\Phi_{22}\bar\Phi_{22} \ ) \ {\bigg]} \qquad, 
\label{dmodinv} 
\eea
as also apparent when the same expression is written in terms of
``characters''  (see Appendix A)
\bea
{\cal T}_{\it{d}} \ = \ {1 \over 3} 
&\!\!\!\!{\bigg[}&\!\!\!\! \Phi_{00}\bar\Phi_{00} \, 
\Lambda' \ {\bigg]} + \chi_{00}\bar\chi_{00} + \chi_{01}\bar\chi_{01} +
\chi_{02}\bar\chi_{02}\ + \ 3 \ ( \ \chi_{10}\bar\chi_{10} +
\chi_{11}\bar\chi_{11} + \chi_{12}\bar\chi_{12} \ ) \nonumber \\   
&\!\!\!\!+&\!\!\!\! 3 \ ( \
\chi_{20}\bar\chi_{20} + \chi_{21}\bar\chi_{21} + \chi_{22}\bar\chi_{22} \ ) 
\qquad ,
\label{dmodinvchar} 
\eea
where the prime denotes a lattice sum without the zero-mode contribution. The
crucial point is now to understand how T-duality \cite{tdual}, that is a symmetry
of the toroidal partition function, can be combined with $\Omega$ on the 
lattice.  Typically, this can
be done using a reflection with respect to some symmetry axis of
the orbifold \cite{bgk}, and in the $Z_3$  case one has two choices.  The first 
corresponds to a reflection ${\cal{I}}_1$ with respect to the 
diagonal of the  unit
cell (or, equivalently, with respect to the vertical axis) and leaves invariant
all the three fixed points.  The second, ${\cal{I}}_2$, is a reflection with
respect to the horizontal axis, and leaves invariant only the origin.  The
two choices are connected to different resolutions of ``fixed-point 
ambiguities'' \cite{fixpo} and give rise to different modular invariants, and
thus to different Klein-bottle projections.  In ref. \cite{ab} the
reflection ${\cal{I}}_2$ was related to ${\cal{I}}_1$ on a
lattice rotated by an angle $\pi/6$.  If the T-duality is associated to 
${\cal{I}}_1$, the $\Omega$ projection fixes states that satisfy the
$\bf{p}=\bar{\tilde{\bf{p}}}$ condition, equivalent to a slice of the 
lattice with
$n_2=0,m_1=0$, with conventions chosen in such a way that
the complex coordinate $z$ corresponds to $i (x_1 + i x_2)$.  The closed
twist produces the amplitudes $(\Phi_{g,h-h'} \, \delta_{g,g'})$ from the
bulk term $(\Phi_{g,h} \, \bar\Phi_{g',h'})$, and the resulting
Klein-bottle amplitude is    
\be  
{\cal K} = {1\over 2} \ \left[ \
\sum_{m,n} \,  { ( \ e^{-2 \pi t} \ )^{{1 \over 2 X_2 Y_2} [m^2 +  {Y_2}^2 n^2]}
\over \eta^2 (2 i t)} \ + \ 3 \ \Phi_{10} \ + \ 3 \ \Phi_{20} \ \right]
\qquad .  \label{kleindirc2} 
\ee
Notice that eq. (\ref{kleindirc2}) exactly symmetrizes the ${\cal T}_{\it{d}}$ 
amplitude, 
as neatly evidentiated by the 
corresponding expression in terms of
characters 
\bea  
{\cal K} &=&  {1\over 2} \ {\bigg[} \ {\sum_{m,n}}' \,  
{ ( \ e^{-2 \pi t} \ )^{{1 \over
2 X_2 Y_2} [m^2 + {Y_2}^2 n^2]} \over \eta^2 (2 i t)} \, + 
\chi_{00} + \chi_{01} +
\chi_{02} \nonumber \\ &+& 3 \
( \ \chi_{10} + \chi_{11} + \chi_{12} \ ) + \ 3 \ ( \ \chi_{20} + \chi_{21} +
\chi_{22} \ )\, {\bigg]} \qquad,  
\label{kdirchar} 
\eea 
where again the primed sum means that we
are extracting the term corresponding to $m=n=0$.  An $S$ modular transformation
exposes the tree channel  
\be 
\tilde{\cal K} = {2^{D/2}\over 2} \ \left[ \ 2 X_2
\ \sum_{m,n} \,  { ( \ e^{-2 \pi \ell} \ )^{{X_2 \over Y_2} [m^2 + 
{Y_2}^2 n^2]} \over \eta^2
(i \ell)} \ + \ \sqrt{3} \ \Phi_{01} \ + \ \sqrt{3} \ \Phi_{02} \ \right]
\qquad , 
\label{kleintrac2}
\ee
where, as usual, the powers of two account for additional dimensions and
come from the (omitted) modular measure.  Because of 
T-duality, the lattice sum in (\ref{kleindirc2}) contains both momenta and
windings.  Consequently, the transverse channel depends on a ratio of volumes, 
and is independent of the radius $R$.  The consistency of (\ref{kleintrac2})
is precisely linked to the value of $X_2$.  It
generalizes to the irrational case the property, familiar from Rational
Conformal Field Theory,  that only the identity appears in the transverse
channel of diagonal invariants.  The
``massless contribution'' is indeed  
\be  
\tilde{\cal K}_0 = {2^{D/2}\over 2} \ 
3 \sqrt{3} \ \chi_{00} \qquad ,  
\label{ktram0c2} 
\ee
as expected from the complete projector in the tree channel.

To construct the annulus amplitude, two observations are in order.
First, the transverse annulus must contain only the states that can be
reflected from a boundary or, equivalently, that are paired with their 
conjugates in the bulk GSO.  This amounts to selecting terms with $\bf{p}=
- \bar{\tilde{\bf{p}}}$.  From (\ref{momenta}) and the form of 
${\cal{I}}_2$, it is
easy to see \cite{ab} that only $m_1$ and $n_2$ survive the projection, but with
the additional condition that both $2 X_1 m_1$ and $2 X_1 n_2$ must be even. 
Second, the Chan Paton matrices reflect the structure of the orbifold group
\cite{ps}, and in the $Z_3$ case one has 
\be
Tr \, [ \, A_k \, ] \ = \  N \, + \omega^k
\, M \, + \, \bar{\omega}^k \, {\bar M} \quad .
\label{z3matrix}
\ee
In the diagonal annulus amplitude, however, we expect both untwisted and
twisted chiral blocks, {\it without} projections.  In
other words, the traces of the breaking matrices $A_k$ with $k \ne 0$ should
vanish, allowing  only a single Chan Paton charge.  Introducing a suitable
projector, the transverse channel annulus becomes 
\bea 
\tilde{\cal A} &\!\!\!=&\!\!\! {2^{- D/2}\over 2} \, {N^2 \over 2} \, 
\biggl[ \,
{X_2 \over 2} \, \sum_{\epsilon_1 , \epsilon_2=0,1} \, \sum_{m,n} \,  { ( \
e^{-2 \pi \ell} \, )^{{X_2 \over 4 Y_2} [m^2 +  {Y_2}^2 n^2]} \, e^{2 \pi i X_1
( m \epsilon_1 + n \epsilon_2)} \over \eta^2 (i \ell)}  \nonumber\\
&\!\!\!+&\!\!\!   \sqrt{3} \,
\Phi_{01} \, + \, \sqrt{3} \, \Phi_{02} \, \biggr] \qquad , 
\label{atrac2}
\eea
where the completeness of the projector is due to the zero mode contributions
that supply exactly the necessary factor of $4$ in front of the $\eta^{-2}$
term.  The ``massless contribution'' is then
\be
\tilde{\cal A}_0 = 2^{-D/2} \ {N^2 \over 2} \ 3 \sqrt{3} \ \chi_{00}
\qquad . 
\label{atram0c2}
\ee
As usual, the direct channel annulus amplitude exhibits windings and
momenta shifted by the (quantized) value of $X_1$ 
\be
{\cal A} = {N^2 \over 2} \ \left[ \, \sum_{\epsilon_1,\epsilon_2=0,1} \,
\sum_{m,n} \,  { ( \ e^{-2 \pi t} \ )^{{1 \over 2 X_2 Y_2} [ (m+X_1 \epsilon_1)^2
+  {Y_2}^2 (n+X_1 \epsilon_2)^2 ]} \over \eta^2
({i t \over 2})} \, + \, 3 \, \Phi_{10} \, + \, 3 \, \Phi_{20} \, \right]
\, , 
\label{adirc2}
\ee
and in terms of characters is
\bea
{\cal A} = {N^2 \over 2} &\!\!\!\!{\bigg[}&\!\!\!\! \chi_{00} + \chi_{01} + \chi_{02} + 
\sum_{\epsilon_1,\epsilon_2=0,1} \, {\sum_{m,n}}' \,  { ( \ e^{-2 \pi t} \ )^{{1
\over 2 X_2 Y_2} [ (m+X_1 \epsilon_1)^2 +  {Y_2}^2 (n+X_1 \epsilon_2)^2 ]} \over
\eta^2 ({i t \over 2})} \nonumber \\
+ \, 3 &\!\!\!(&\!\!\! \chi_{10} + \chi_{11} + \chi_{12} \ ) \, + \, 3 \, ( \,
\chi_{20} + \chi_{21} + \chi_{22} \, ) \, \bigg] \,  \qquad.
\label{adirchar}
\eea
Sector by sector, the transverse M\"obius amplitude is the geometric 
mean of the transverse annulus
and Klein bottle amplitudes, with signs needed to normalize correctly 
the projector \cite{toroidal,zurB,carlo,ab} 
\bea 
\tilde{\cal M} = 2 \, {N \over 2} \, {e^{i \pi \over 12}} &\!\!\!\!{\bigg[}&\!\!\!\!
\, X_2 \, \sigma_{00} \, \sum_{\epsilon_1 , \epsilon_2=0,1} \, \sum_{m,n} \,
{\gamma_{\epsilon_1,\epsilon_2} ( \ e^{-2 \pi \ell} \, )^{{X_2 \over Y_2} [m^2
+  {Y_2}^2 n^2]} \, e^{2 \pi i X_1 ( m \epsilon_1 + n \epsilon_2)} \over \eta^2
(i \ell + {1 \over 2})} \nonumber \\ 
&\!\!\!\!+&\!\!\!\! \sigma_{01} \sqrt{3} \, \Phi_{01}
\, + \, \sigma_{02}\sqrt{3} \, \Phi_{02} \, \bigg] \qquad .  
\label{moetrac2}
\eea
Notice that, barring the overall phase that appear in the definition 
of hatted
quantities (see Appendix A), a sensible projector requires 
$\sigma_{00}=\sigma_{01}=\sigma_{02}$ and the condition 
\be
\sum_{\epsilon_1 , \epsilon_2=0,1} \ \gamma_{\epsilon_1,\epsilon_2} \ = \ 2
\qquad , 
\label{segni}
\ee
in such a way that the ``massless'' term includes only the
``hatted'' identity: 
\be
\tilde{\cal M}_0 = {2 N \over 2} \sigma_{00} \ 3 \sqrt{3} \ \hat\chi_{00}
\qquad .
\ee
It should be appreciated that the transverse ``massless'' terms of
$\tilde{\cal K}$, $\tilde{\cal A}$ and $\tilde{\cal M}$ combine to give a
perfect square.  It is also nice that the three amplitudes must contribute
the {\it same} coefficient, so that the size of the Chan-Paton group in a
critical model would depend solely on the additional coordinates, with a rank
reduction by a factor of two, as already observed in \cite{ab}.  As noticed
before, the reflection coefficients in front of boundaries and crosscaps are
independent of the radius $R$.  In Type I vacua the sign  $\sigma_{00}$ would be
fixed by the cancellation of the identity tadpole, as we shall see in the next
Section.  The matrix $P^{-1}$ allows one to display the direct channel M\"obius
amplitude  
\bea 
{\cal M} &\!\!\!=&\!\!\! {N\over 2} \, {e^{i \pi \over 12}} \, 
\sigma_{00} \, {\bigg[} \,
\sum_{\epsilon_1,\epsilon_2=0,1} \, \sum_{m,n} \, 
{\gamma_{\epsilon_1,\epsilon_2} ( \ e^{-2 \pi t} \ )^{{1 \over 2 X_2 Y_2} [
(m+X_1 \epsilon_1)^2 +  {Y_2}^2 (n+X_1 \epsilon_2)^2 ]} \over \eta^2 ({i t
\over 2} + {1 \over 2})} \nonumber\\
&\!\!\!+&\!\!\! \, 3 \, {e^{-{i \pi \over 9}}} \,
(\Phi_{11} \, + \, \Phi_{22} \,)\, {\bigg]} \qquad ,   
\label{mdirc2} 
\eea 
an expression that, in terms of hatted characters, 
\bea 
{\cal M} = {N \over 2} \, \sigma_{00} 
&\!\!\!{\bigg[}&\!\!\! \sum_{\epsilon_1,\epsilon_2=0,1} \,
{\sum_{m,n}}' \,  {\gamma_{\epsilon_1,\epsilon_2} \,( \ e^{-2 \pi t} \ )^{{1
\over 2 X_2 Y_2} [ (m+X_1 \epsilon_1)^2 +  {Y_2}^2 (n+X_1 \epsilon_2)^2 ]} \over
\hat{\eta}^2 ({i t \over 2} + {1 \over 2})} \nonumber \\ + 
\gamma_{00} &\!\!\!(&\!\!\! \hat\chi_{00} +
\hat\chi_{01} + \hat\chi_{02} \, ) \, +  \, 3 \, ( 
\, \hat\chi_{10} - \hat\chi_{11} - \hat\chi_{12} \ ) \, + \, 3 \, ( \,
\hat\chi_{20} - \hat\chi_{21} - \hat\chi_{22} \, ) \, {\bigg]} ,  
\label{mdirchar} 
\eea 
is manifestly compatible with the annulus amplitude of eq. (\ref{adirchar}) 
and allows both orthogonal and symplectic global Chan-Paton groups.

It is also interesting to
look at ${\cal{I}}_2$, the reflection  with respect to
the horizontal axis.  In this case, states satisfying the condition  $\bf{p}= -
\bar{\tilde{\bf{p}}}$ flow in the direct Klein bottle amplitude, and only one of
the chiral twist fields survives the unoriented projection, reflecting a different
resolution of the fixed-point ambiguity in the twisted sectors.  In fact, 
\be 
{\cal K} = {1
\over 2} \ \left[ \,{1 \over 4} \,  \sum_{\epsilon_1 , \epsilon_2=0,1} \,
\sum_{m,n} \,  { ( \ e^{-2 \pi t} \, )^{{X_2 \over 2 Y_2} [m^2 +  {Y_2}^2 n^2]}
\, e^{2 \pi i X_1 ( m \epsilon_1 + n \epsilon_2)} \over \eta^2 (2 i t)} \, + \,
\Phi_{01} \, + \, \Phi_{02} \, \right] \, ,   \label{k2dirc2}
\ee 
that in terms of characters becomes
\bea
{\cal K} &\!\!\!=&\!\!\! {1\over 2} \ {\bigg[} \ {1 \over 4}
\,  \sum_{\epsilon_1 , \epsilon_2=0,1} \, {\sum_{m,n}}' \,  { ( \ e^{-2 \pi t} \,
)^{{X_2 \over 2 Y_2} [m^2 +  {Y_2}^2 n^2]} \, e^{2 \pi i X_1 ( m \epsilon_1 +
n \epsilon_2)} \over \eta^2 (2 i t)} \, + \, 
\chi_{00} + \chi_{01} + \chi_{02}  \nonumber \\
&\!\!\!+&\!\!\! \ ( \ \chi_{10} + \chi_{11} + \chi_{12} \ ) + \ ( \
\chi_{20} + \chi_{21} + \chi_{22} \ ) \ {\bigg]} \qquad ,
\label{k2dirchar}
\eea
and it is evident, comparing with (\ref{dmodinvchar}), that one of the three
degenerate twist fields is diagonal while the remaining two are off-diagonal in
the would-be resolved torus amplitude.
In (\ref{k2dirc2}), the projector is at work in the direct channel. 
Consequently, even if in the transverse channel windings and momenta are now
shifted  
\bea
\tilde{\cal K} &\!\!\!=&\!\!\! {2^{D/2}\over 2} \ {\bigg[} {2 \over {4 X_2}} \,
\sum_{\epsilon_1,\epsilon_2=0,1} \, \sum_{m,n} \,  { ( \ e^{-2 \pi \ell} \ )^{{1
\over X_2 Y_2} [ (m+X_1 \epsilon_1)^2 +  {Y_2}^2 (n+X_1 \epsilon_2)^2 ]} \over
\eta^2 (i \ell)} \nonumber\\ 
&\!\!\!+&\!\!\! \, {1 \over \sqrt{3}} \, \Phi_{01} \, + \, {1 \over
\sqrt{3}} \, \Phi_{20} \, {\bigg]} \qquad , 
\label{k2trac2}
\eea
the ``massless contribution'' involves again only the identity 
\be
\tilde{\cal K}_0 = {2^{D/2}\over 2} \  {3 \over \sqrt{3}} \ \chi_{00}
\qquad ,
\label{k2tram0c2}
\ee
with a different reflection coefficient for the crosscap.
The transverse annulus amplitude now receives contributions from states with 
$\bf{p}=\bar{\tilde{\bf{p}}}$, without the need of any projections
\be
\tilde{\cal A} = {2^{-D/2}\over 2} \ N^2 \ {\bigg[} {1 \over 2 X_2}  \ \sum_{m,n} \, 
{ ( \ e^{-2 \pi \ell} \ )^{{1 \over 4 X_2 Y_2} [m^2 + 
{Y_2}^2 n^2]} \over \eta^2
(i \ell)} \ + {1 \over \sqrt{3}} \, \Phi_{01} \, + \, {1 \over
\sqrt{3}} \, \Phi_{20} \, {\bigg]} \qquad .
\label{a2trac2}
\ee
At ``zero mass'', eq. (\ref{a2trac2}) gives 
\be
\tilde{\cal A}_0 = 2^{-D/2} \ {N^2 \over 2} \ {3 \over \sqrt{3}} \ \chi_{00}
\qquad , 
\label{a2tram0c2}
\ee
while an $S$ modular transformation yields the direct channel 
amplitude 
\be
{\cal A} = {N^2 \over 2} \ \left[ \ \sum_{m,n} \, 
{ ( \ e^{-2 \pi t} \ )^{{X_2 \over 2 Y_2} [m^2 + 
{Y_2}^2 n^2]} \over \eta^2
({i t \over 2})} \ + \ \Phi_{10} \ + \ \Phi_{20} \
\right] \qquad , 
\label{a2dirc2}
\ee
or, in terms of characters, 
\bea
{\cal A} &\!\!\!=&\!\!\! {N^2 \over 2} \ {\bigg[} \ \chi_{00} + \chi_{01} + \chi_{02} +
{\sum_{m,n}}' \,  { ( \ e^{-2 \pi t} \ )^{{X_2 \over 2 Y_2} [m^2 + 
{Y_2}^2 n^2]} \over \eta^2
({i t \over 2})} \nonumber \\
&\!\!\!+&\!\!\! \ ( \ \chi_{10} + \chi_{11} + \chi_{12} \ ) + \ ( \
\chi_{20} + \chi_{21} + \chi_{22} \ ) \ {\bigg]} \qquad .
\label{a2dirchar}
\eea
Notice that, due to the quantization of $X_1$, the transverse M\"obius
amplitude involves shifts as well.  The result is
\bea
\tilde{\cal M} &\!\!\!=&\!\!\! 2 \, {N \over 2} \, {e^{i \pi \over 12}} 
\, {\bigg[} \, {1 \over 2
X_2} \, \sum_{\epsilon_1 , \epsilon_2=0,1} \, \sum_{m,n} 
\, {\gamma_{\epsilon_1,\epsilon_2} \,  ( \ e^{-2
\pi \ell} \ )^{{1 \over X_2 Y_2} [ (m+X_1 \epsilon_1)^2 +  {Y_2}^2 (n+X_1
\epsilon_{2})^2 ]} \over \eta^2 (i \ell + {1 \over 2})} \nonumber\\ 
&\!\!\!+&\!\!\! \, {\sigma_{01} \over
\sqrt{3}} \, \Phi_{01} \, + \, {\sigma_{02}\over \sqrt{3}} \, \Phi_{02}
\,  {\bigg]} \qquad ,    
\label{moe2trac2} 
\eea
and consistency requires $\gamma_{00}=\sigma_{01}=\sigma_{02}$ to produce the
correct ``massless'' contribution
\be
\tilde{\cal M}_0 ={2 N \over 2}\, \gamma_{00} \, {3  \over \sqrt{3}} \
\hat\chi_{00} \qquad .
\ee
In the direct channel, the
$\epsilon$-dependence is again crucial.  In fact,
\bea
{\cal M} = {N\over 2} \, {e^{i \pi \over 
12}} &\!\!\!\!{\bigg[}&\!\!\!\! \, {1 \over 2} \, \sum_{\epsilon_1 , \epsilon_2=0,1} \, \sum_{m,n} \, 
{ \gamma_{\epsilon_1,\epsilon_2}\, ( \ e^{-2 \pi t} \ )^{{X_2 \over 2 Y_2} [m^2 + 
{Y_2}^2 n^2]} \, e^{2 \pi i X_1 ( m \epsilon_1 + n \epsilon_2)} \over \eta^2
({i t \over 2} + {1 \over 2})} \nonumber\\
+ \, {e^{-{i \pi \over 9}}} \, \gamma_{00} &\!\!\!(&\!\!\! \Phi_{11}
\, + \, \Phi_{22} \, ) \ {\bigg]} \qquad , 
\label{m2dirc2}
\eea
and the correct normalization of the projector demands that 
\be
\sum_{\epsilon_1 , \epsilon_2=0,1} \ \gamma_{\epsilon_1,\epsilon_2} \ = \ 
2 \ \delta
\label{segnibis}
\ee
be satisfied, with $\delta$ a sign, in such a way 
that, in terms of hatted characters, 
\bea
{\cal M} = {N \over 2} \, &\!\!\!\!{\bigg[}&\!\!\!\! \, \delta \, ( 
\, \hat\chi_{00} + \hat\chi_{01} +
\hat\chi_{02} \, ) +  \sum_{\epsilon_1 , \epsilon_2=0,1} \, {\sum_{m,n}}' \, 
{ ( \ e^{-2 \pi t} \ )^{{X_2 \over 2 Y_2} [m^2 + 
{Y_2}^2 n^2]} \, e^{2 \pi i X_1 ( m \epsilon_1 + n 
\epsilon_2)} \over \hat{\eta}^2
({i t \over 2} + {1 \over 2})}  \nonumber \\ 
+ \, \gamma_{00} &\!\!\!(&\!\!\! \hat\chi_{10} - \hat\chi_{11} - 
\hat\chi_{12} \,) \, + \, 
\gamma_{00} \, (\, \hat\chi_{20} -
\hat\chi_{21} - \hat\chi_{22}\,)\, {\bigg]} \,  \qquad.   
\label{m2dirchar}  
\eea
Notice that the sign $\gamma_{00}$ is fixed in Type I vacua by the cancellation 
of the $\chi_{00}$ tadpole while $\delta$ determines the type 
of the Chan-Paton group,
whose rank is reduced by a factor of $2$ as a result of 
the equality of the reflection coefficients for boundaries and crosscaps.

\section{Type I vacua in six and four dimensions}

We now apply the results of the previous Section to the construction of
Type I vacua in six and four dimensions, starting from the corresponding diagonal
$Z_3$ orbifold of the Type IIB theory.  In $D=10-d$ ($D=4$ or $6$), one can
resolve the fixed-point ambiguity in several ways, allowing 
a number $n=\prod_{i=1}^{d/2} k_i$ (with $k_i$ equal to $1$ or $3$) of 
twist fields associated to as many fixed
points in the direct Klein bottle amplitude.  We thus find
three models with $n$ equal to $1$, $3$ and $9$ in $D=6$, and four models with
$n$ equal to $1$, $3$, $9$ and $27$ in $D=4$.  Using supersymmetric chiral
blocks (see Appendix A), the Type IIB diagonal modular invariant is  
\bea
{\cal T_{\it{d}}} &\!\!\!=&\!\!\! {1 \over 3} \, {\bigg[} \ \rho_{00} \,
\bar\rho_{00} \  {\prod_{i=1}^{d/2} \Lambda_{i}}  + \rho_{01}\bar\rho_{01} + 
\rho_{02}\bar\rho_{02} + \ 3^{d/2} \ ( \ \rho_{10}\bar\rho_{10} + \rho_{11}\bar\rho_{11} 
+ \rho_{12}\bar\rho_{12} \ )  \nonumber \\  
&\!\!\!+&\!\!\! 3^{d/2} \ ( \ \rho_{20}\bar\rho_{20} + \rho_{21}\bar\rho_{21}
+  \rho_{22}\bar\rho_{22} \ ) \ {\bigg]} \qquad,
\label{dd6modinv}
\eea
that in terms of supersymmetric characters becomes
\bea
{\cal T}_{\it{d}} \ &\!\!\!=&\!\!\! \ {1 \over 3} \ {\bigg[} \ \rho_{00}\bar\rho_{00} 
\, (\, {\prod_{i=1}^{d/2} \Lambda_{i}} \, )' \, {\bigg]} + \chi_{00}\bar\chi_{00} + 
\chi_{01}\bar\chi_{01} + \chi_{02} \bar\chi_{02} \nonumber  \\     
&\!\!\!+&\!\!\! 3^{d/2} \ ( \ \chi_{10}\bar\chi_{10} + \chi_{11}\bar\chi_{11} + 
\chi_{12}\bar\chi_{12} \ ) \, +\, 3^{d/2} \, ( \, 
\chi_{20}\bar\chi_{20} + \chi_{21}\bar\chi_{21} + \chi_{22}\bar\chi_{22} \, ) 
\ . 
\label{dd6modinvchar} 
\eea
From now on, we do not specify the explicit form of the lattice sums, 
that should be clear from the previous Section.  In complete analogy with the
$c=2$ case 
\be
{\cal K} = {1\over 2} \ \left[ \ \rho_{00} \ {\prod_{i=1}^{d/2} 
\Gamma_{i}^{K}} + \ n \ 
\rho_{10} \ + \ n\ \rho_{20} \ \right] \qquad ,
\label{kd6dir}
\ee
and extracting the zero modes from the lattice sums, it can be written in terms
of $Z_3$-characters as 
\bea
{\cal K} &\!\!\!=&\!\!\! {1\over 2} \ [ \, \rho_{00} \
(\, {\prod_{i=1}^{d/2}  \Gamma_{i}^{K} \, )'} \, + \,
\chi_{00} + \chi_{01} + \chi_{02} +  \nonumber \\ &\!\!\!+&\!\!\! \ n \ ( \
\chi_{10} + \chi_{11} + \chi_{12} \ ) + \ n\ ( \ \chi_{20} + \chi_{21} +
\chi_{22} \ ) \, ]  \qquad . 
\label{kd6char} 
\eea
It is clear from (\ref{kd6char}) that only the twist fields 
associated to $n$ of the $3^{d/2}$ fixed points survive the
unoriented projection.  Exposing the transverse channel by an $S$
transformation gives
\bea
\tilde{\cal K} &=& {2^{D/2}\over 2} \ \left[ \, n \, 3^{-d/2} \, ( \rho_{00} \,
{\prod_{i=1}^{d/2}  \tilde{\Gamma}_{i}^{K}} \, + \, \rho_{01} \, + \, \rho_{02} \,
) \,  \right] \nonumber\\ 
&=& {2^{D/2}\over  2} \, n \, 3^{(1-d/2)} \, \left[ \ \chi_{00}
+ {1 \over 3} \, \rho_{00} \
(\, {\prod_{i=1,d/2}  \tilde{\Gamma}_{i}^{K}} \, )' \right] \qquad ,
\label{kd6tra} 
\eea
and it is worthwhile to observe that, as in the $c=2$ model, the complete
projector is restored in virtue of the right values of the $X_2$ moduli in
the compactified directions.  A single type of Chan-Paton charge is allowed in
the annulus amplitude, that can be associated to Dirichlet $(9-d/2)$-branes at
angles \cite{bgk}, the natural objects in the theory after $d/2$ T-duality
transformations.  Thus  
\be 
{\cal A} = {N^{2}\over 2} \ \left[ \
\rho_{00} \ {\prod_{i=1}^{d/2} 
\Gamma_{i}^{A}} + \ n \  \rho_{10} \ + \ n\ \rho_{20} \
\right] \qquad , \label{ad6dir}
\ee
and, as should be clear at this point, the transverse channel gives back the
right contributions 
\bea
\tilde{\cal A} &\!\!\!=&\!\!\! {2^{-D/2}\over 2} \ N^{2} \left[ \, n \, 3^{-d/2} \, (
\rho_{00} \, {\prod_{i=1}^{d/2}  \tilde{\Gamma}_{i}^{A}} \, + \, \rho_{01} \, + \,
\rho_{02} \, ) \,  \right] \nonumber\\
&\!\!\!=&\!\!\! {2^{-D/2}\over 
2} \ N^{2} \ n \, 3^{(1-d/2)} \, \left[ \ \chi_{00}
+ {1 \over 3} \, \rho_{00} \
(\, {\prod_{i=1}^{d/2}  \tilde{\Gamma}_{i}^{A}} \, )' \right] \qquad .
\label{ad6tra}
\eea
The M\"obius amplitude completes the one-loop Type I partition function.  In
the open-string loop channel, 
\be
{\cal M} = - (-1)^{d/2} {N\over 2} \ \left[ \, \rho_{00} \ {\prod_{i=1}^{d/2} 
\Gamma_{i}^{M}}
+ \, n \,  \rho_{11} \, + \, n \, \rho_{22} \, \right] \qquad ,
\label{md6dir}
\ee
while in the transverse channel 
\bea
\tilde{\cal M} &\!\!\!=&\!\!\! - 2 \, {N\over 2} \, n \, 3^{-d/2} \, \left[ \, 
\rho_{00} \, {\prod_{i=1}^{d/2}  \tilde{\Gamma}_{i}^{M}} \, + \, \rho_{01} \, + \,
\rho_{02} \,  \right] \nonumber\\ 
&\!\!\!=&\!\!\! - 2 \, {N \over 2}  \, n \, 3^{(1-d/2)} \,
\left[ \, \hat\chi_{00} + {1 \over 3} \, \rho_{00} \
(\, {\prod_{i=1}^{d/2}  \tilde{\Gamma}_{i}^{M}} \, )' \, \right] 
\qquad . 
\label{md6tra} 
\eea
Notice that, in this case, the sign is fixed by the tadpole of $\chi_{00}$.  From
$\tilde{\cal{K}}$, $\tilde{\cal{A}}$ and $\tilde{\cal{M}}$ we get
$N=2^{D/2}$, with symmetrization of the Chan-Paton charges for the gauge 
vectors in four dimensions and
antisymmetrization in six dimensions.  As a consequence, all the $D=6$
models have an $SO(8)$ gfe group, while all the $D=4$ models have
an $Sp(4)$ gauge group.  Their massless spectra are summarized in Table
1 and in Table 2. 
\begin{table}
\begin{center}
\begin{tabular}{|c|c|c|c|c|} 
\hline
\, & closed & closed & open & open  \\
\hline 
$n$ & T & H & CP group & H\\
\hline
\hline
$1$ & 8 & 13 & $SO(8)$ & 2 \, $({\bf{28}})$\\
\hline
$3$ & 6 & 15 & $SO(8)$ & 4 \,$({\bf{28}})$\\
\hline
$9$ & 0 & 21 & $SO(8)$ & 10 \,$({\bf{28}})$\\
\hline
\end{tabular}
\end{center}
\caption{Massless spectra of $D=6$ models.}
\label{tab1}
\end{table}

The closed oriented massless spectrum of Type IIB on the $Z_3$
orbifold in $D=6$ results in $N=(2,0)$ supergravity coupled to $21$ tensor
multiplets.  The unoriented truncation produces $N=(1,0)$ models with $12+n$
hypermultiplets and $9-n$ tensor multiplets.  The open sector adds the gauge
multiplet and $n+1$ hypermultiplets in the adjoint representation of
$SO(8)$.  These models were already described in \cite{ab}, and as
open descendants of Gepner models in \cite{gepner}.  The model with $n=3$
also corresponds to a recently described open descendant of a $Z_{3L} \times
Z_{3R}$ asymmetric orbifold
\cite{noi}.  It is amusing to stress how all these descriptions are
related to different resolutions
of the fixed-point ambiguities in the diagonal modular
invariant.  It is also easy to check that the gravitational
anomaly cancellation condition
\be
H\, - \, V \ = \ 273 \, - 29 \, T  
\label{anom}
\ee
is satisfied for every integer value of $n$ between $1$ and $9$.  As 
usual, the models with several tensor multiplets are anomaly-free due 
to a GSS mechanism \cite{gs,aug}.   

The four dimensional models are related to a compactification of the Type IIB
on a Calabi-Yau threefold with Hodge numbers $h_{1,1}=0$ and $h_{1,2}=36$. 
After the unoriented truncation, the massless closed spectrum contains the $N=1$
supergravity multiplet coupled to $V = {(27-n)/2}$ vector multiplets and to $10
+ n + V$ chiral multiplets.  A gauge vector together with $n+3$ chiral
multiplets in the adjoint of the $Sp(4)$ Chan-Paton groups results from the open
and unoriented massless spectrum.  These models, differently from the geometric
open descendants \cite{chiral}, are not chiral and are clearly anomaly-free. 
Compared to the model with
$n=27$ (``the true diagonal''), the presence of additional 
vector multiplets in the
unoriented closed spectrum reduces the number of marginal deformations.
\begin{table}
\begin{center}
\begin{tabular}{|c|c|c|c|c|} 
\hline
\, & closed & closed & open & open  \\
\hline 
$n$ & C & V & CP group & C\\
\hline
\hline
$1$ & 24 & 13 & $Sp(4)$ & 4 \, $({\bf{10}})$\\
\hline
$3$ & 25 & 12 & $Sp(4)$ & 6 \, $({\bf{10}})$\\
\hline
$9$ & 28 & 9 & $Sp(4)$ & 12 \, $({\bf{10}})$\\
\hline
$27$ & 37 & 0 & $Sp(4)$ & 30 \, $({\bf{10}})$\\
\hline
\end{tabular}
\end{center}
\caption{Massless spectra of $D=4$ models.}
\label{tab2}
\end{table}
 
\section{Conclusions and discussion}

Starting from a pair of free bosonic fields propagating 
on a $Z_3$ orbifold of a two-torus, we have discussed the open descendants of
the diagonal model.  Its consistency, as
stressed in ref. \cite{ab}, is deeply connected to discrete values of
some geometric moduli.  In fact, the diagonal GSO projection results from 
the combination of the geometric orbifold action and a 
T-duality.  For the $Z_3$
group, on a two-dimensional lattice there are two possible choices, corresponding
to two different resolutions of the fixed-point ambiguities, that give rise to 
different classes of open descendants.
We have then discussed the application to (already
known)  six dimensional and to (new) four dimensional open descendants of
Type IIB $Z_3$ diagonal orbifolds.  Their spectra are parameterized by the
number $n$ of fixed points surviving the closed unoriented projection.  In
$D=6$, $n$ can take the values $1$, $3$ and $9$, giving rise to Type I models
with an $SO(8)$ Chan-Paton gauge group and $13 + 2 n$ hypermultiplets and $9-n$
tensor multiplets.  In $D=4$, $n$ can be $1$, $3$, $9$ and $27$, yielding
non-chiral Type I vacua with an $Sp(4)$ Chan-Paton gauge group, $V=(27 - n)/2$
additional vector multiplets and $13 + 2n + V$ chiral
multiplets.  An inspection of the complete one-loop partition function and of
the Rational Conformal Field Theory examples suggests that all these models
should be consistent {\it for every integer value} of
$n$ between $1$ and $9$ in six dimensions and {\it for every 
odd-integer value} of
$n$ between $1$ and $27$ in four dimensions.  In order to implement this 
conjecture, one should be able to find a suitable
involution that represents the combined action of $\Omega$ and the 
T-duality on the lattice and leaves 
exactly
$n$ fixed points invariant.  This is not allowed for orbifolds that are products
of two-tori, but more complicated slices of four dimensional or
six-dimensional lattices could exist that realize this settings.  

It would be interesting to extend this construction to open descendants of
$Z_N$ orbifolds, and to investigate the possibility of introducing anti-branes in
this context.  Notice that we do not find, for the $Z_{3}$ case, unsolvable tadpole 
conditions as in recently proposed open descendants of the Type IIA in 
four dimensions \cite{twoa}.  
Finally, it should be noticed that all these models are defined perturbatively
and are tightly constrained by Conformal Field Theory consistency conditions on
surfaces with boundaries and/or crosscaps.  This is true, in 
particular, for the six
dimensional model with zero tensor multiplets.  One is not allowed to append to
a given closed unoriented spectrum an open sector that does not respect the
aforementioned constraints.  Some recently 
proposed ``non-perturbative'' orientifolds are built 
violating the anomaly-cancellation conditions and calling for non-perturbative
states that supply the missing multiplets \cite{kaku}.  We have no way to check
the consistency of these models, but certainly we are able to define
perturbative orientifolds compatibly with the closed unoriented spectra of some
of the models of ref. \cite{kaku}.  

\section{Acknowledgments}

It is a pleasure to thank 
C.~Angelantonj, M.~Bianchi, J.F.~Morales, A.~Sagnotti, and
Ya.S.~Stanev for interesting discussions and C. Schweigert for e-mail
correspondence. While this work was being completed, Carlo Angelantonj
informed us that Blumenhagen and collaborators were considering related
issues \cite{bnew}.

\section{Note added}

For the sake of comparison with \cite{bnew}, where models slightly 
different from these are presented, we would like to add the following 
comments.  Actually, there are more solutions, both in six and in four 
dimensions, connected to the presence of discrete open-string 
Wilson-lines \cite{bs,toroidal,carlo}.  
\begin{table}
\begin{center}
\begin{tabular}{|c|c|c|c|c|c|} 
\hline
\, & closed & closed & open & open & open \\
\hline 
$n$ & T & H & CP group & H & H \\
\hline
\hline
$1$ & 8 & 13 & $Sp(8)$ & 1 \, $({\bf{36}})$ & 1 \, $({\bf{28}})$\\
\hline
$3$ & 6 & 15 & $Sp(8)$ & 1 \, $({\bf{36}})$ & 3 \,$({\bf{28}})$\\
\hline
$9$ & 0 & 21 & $Sp(8)$ & 1 \, $({\bf{36}})$ & 9 \,$({\bf{28}})$\\
\hline
\end{tabular}
\end{center}
\caption{Massless spectra of $D=6$ models with $\gamma = -1$.}
\label{tab3}
\end{table}
In fact, the M\"obius-strip amplitude in 
(\ref{md6dir}) should be written as
\be
{\cal M} = - {N\over 2} \ \left[ \, \rho_{00} \ {\prod_{i=1}^{d/2} 
\Gamma_{i}^{M}}
+ \, (-1)^{d/2} \, n \,  \rho_{11} \, + \, (-1)^{d/2} \, n 
\, \rho_{22} \, \right] \qquad ,
\label{mdgen}
\ee
or, in terms of characters,
\bea
{\cal M} = - {N\over 2} \ &\!\!\!{\bigg[}&\!\!\! \, \gamma \, ( \, \hat\chi_{00} 
\, + \, \hat\chi_{01} \, + \, \hat\chi_{02} \, ) \, + \, \rho_{00} 
\, ( \, {\prod_{i=1}^{d/2} 
\Gamma_{i}^{M}} \, )' \nonumber\\
+ \, (-1)^{d/2} \, n \,  &\!\!\!(&\!\!\! 
\, \hat\chi_{10} - \hat\chi_{11} - \hat\chi_{12} \ ) \, + \,  (-1)^{d/2} \, n  \, ( \,
\hat\chi_{20} - \hat\chi_{21} - \hat\chi_{22} \, ) \, {\bigg]} \qquad ,
\label{mdgchar}
\eea
where $\gamma$ is a sign hidden in the lattice sum, as pointed out in 
Section 2.  Comparing 
with the annulus amplitude of eq. (\ref{ad6dir}), one can see 
that $\gamma = +1$ leads to orthogonal Chan-Paton 
groups, while $\gamma = -1$ leads to symplectic ones.  
The six-dimensional models in Table 1 correspond to $\gamma = 
+1$, while the four-dimensional models in Table 2 correspond to $\gamma = 
-1$.  Two other series of models, whose 
massless spectra are reported in Tables 3 and 4, are thus available. 
\begin{table}
\begin{center}
\begin{tabular}{|c|c|c|c|c|c|} 
\hline
\, & closed & closed & open & open & open \\
\hline 
$n$ & C & V & CP group & C & C \\
\hline
\hline
$1$ & 24 & 13 & $SO(4)$ & 3 \, $({\bf{6}})$ & 1 \, $({\bf{10}})$\\
\hline
$3$ & 25 & 12 & $SO(4)$ & 3 \, $({\bf{6}})$ & 3 \, $({\bf{10}})$\\
\hline
$9$ & 28 & 9 & $SO(4)$ & 3 \, $({\bf{6}})$ & 9 \, $({\bf{10}})$\\
\hline
$27$ & 37 & 0 & $SO(4)$ & 3 \, $({\bf{6}})$ & 27 \, $({\bf{10}})$\\
\hline
\end{tabular}
\end{center}
\caption{Massless spectra of $D=4$ models with $\gamma = +1$.}
\label{tab4}
\end{table}
Notice that the 
closed spectra are equal to the previous ones, but the open sectors 
are changed.  In particular, in $D=6$ one has the gauge multiplet and 
one hypermultiplet in the adjoint representation of $Sp(8)$, with $n$ 
hypermultiplets in the antisymmetric representation.  Eq. 
(\ref{anom}) is again satisfied, because the Wilson line affects both 
the vector and a corresponding untwisted hypermultiplet.  The gauge 
anomalies in the models with several tensor multiplets are again 
absent, due to the GGS mechanism \cite{gs,aug}.   

In $D=4$, one 
has the gauge vector multiplet and three chiral multiplets in the adjoint 
representation of $SO(4)$, together with $n$ chiral multiplets in the symmetric 
representation.  All these models are not chiral and thus 
anomaly-free, and coincide with the $Z_{3}$-models in ref. 
\cite{bnew}.

\section{Appendix A: notations and conventions}

Let us start by defining the chiral blocks for the $c=2$ model describing a 
pair of 
free bosons on the $Z_3$
orbifold.  The chiral traces in the untwisted sector are
\be
\Phi_{0,h} \ = \ \left[ \, q^{1/12} \prod_{n=1}^{\infty} \, (\, 
1 \, - \, \omega^{h} q^{n} \, ) \, (\, 1 \, - \, 
{\bar{\omega}}^{h} q^{n} \, ) \, \right]^{-1} \qquad ,
\label{Phi}
\ee
while in the twisted sectors
\be
\Phi_{1,h} \ = \Phi_{2,-h} \ = \left[ \, q^{-{1/36}} \prod_{n=1}^{\infty}
\, (\,  1 \, - \, \omega^{h} q^{n-2/3} \, ) \, (\, 1 \, - \, 
{\bar{\omega}}^{h} q^{n-1/3} \, ) \, \right]^{-1} \qquad ,
\label{Phitw}
\ee
with $\omega=e^{2 \pi i/3}$ and $h=(0,1,2) \, mod \, 3$.  Notice that 
$\Phi_{0,0} = {\eta}^{-2}$.

The chiral supertraces entering the Type II and Type I superstring orbifold
models  are defined by
\be
\rho_{g,h}\equiv {\rm Tr}_{\rm NS,g} \frac{1}{2}(1-(-)^{F})h
q^{L_0-\frac{c}{24}}- {\rm Tr}_{\rm R,g}\frac{1}{2}(1+(-)^{F})h
q^{L_0-\frac{c}{24}} \quad ,
\label{rhogh}
\ee
where $g,h\in Z_3$, the trace runs over the 
$g$-twisted sector with a plus sign for NS states and a minus sign for R  states
and we are omitting the measure and the contribution of non-compact
coordinates.  For the $h$ projection 
in a given $g$-twisted sector, one thus obtains  
\bea
\rho_{00}&\equiv& \frac{1}{2}\sum_{\alpha,\beta= 0,1/2}
(-)^{2\alpha+2\beta+4\alpha\beta}
\ \frac{\vartheta{\alpha \brack \beta}^4}{\eta^{4+d}}\nonumber\\
\rho_{0h}&\equiv&\frac{1}{2}\sum_{\alpha,\beta= 0,1/2}
(-)^{2\alpha+2\beta+4\alpha\beta}
\left(\frac{\vartheta{\alpha \brack \beta}}{\eta}\right)^{4-{d/2}}
 \,\prod_{i= 1}^{d/2} (2sin\pi h_i)\frac{\vartheta{\alpha \brack 
\beta+h_i}}
{\vartheta{\frac{1}{2} \brack \frac{1}{2}+h_i}}~~~h\neq 0\nonumber\\
\rho_{gh}&\equiv&-(i)^{\frac{d}{2}}\frac{1}{2}\sum_{\alpha,\beta= 0,1/2}
(-)^{2\alpha+2\beta+4\alpha\beta}
\left(\frac{\vartheta{\alpha \brack \beta}}{\eta}\right)^{4-{d/2}}
\, \prod_{i= 1}^{d/2} \frac{\vartheta{\alpha+g_i \brack \beta+h_i}}
{\vartheta{\frac{1}{2}+g_i \brack \frac{1}{2}+h_i}}
~~~g,h\neq 0\label{rho} \quad ,
\eea
where ${\vartheta{\alpha \brack \beta}}$ are the standard Jacobi
theta functions with characteristics and
$\sum_{i=1}^{d/2} \, g_i=  \sum_{i=1}^{d/2} \, h_i = 0 \ (\, mod \ 1 \, )$.

Under S-modular transformations ($\tau\rightarrow
-{1}/{\tau}$)
\bea
\Phi_{00} &\rightarrow& (-i \tau)^{-1} \, \Phi_{00}\nonumber\\
\Phi_{0h} &\rightarrow
& (\sqrt{3}) \, \Phi_{h0}~~~h\neq 0\nonumber\\
\Phi_{h0} &\rightarrow
& (1/\sqrt{3}) \, \Phi_{0,-h}~~~h\neq 0\nonumber\\
\Phi_{gg} &\rightarrow&
(e^{i \pi/18}) \, \Phi_{g,-g}~~~g\neq 0\nonumber\\
\Phi_{g,-g} &\rightarrow
& (e^{-{i \pi/18}}) \, \Phi_{-g,-g}~~~g\neq 0 \quad ,
\label{SPhi}
\eea
and
\bea
\rho_{00} &\rightarrow& (-i \tau)^{-d/2} \, \rho_{00}\nonumber\\
\rho_{0h} &\rightarrow
& (2 sin{\pi h})^{\frac{d}{2}}\, \rho_{h0}~~~h\neq 0\nonumber\\
\rho_{h0} &\rightarrow
& (2 sin{\pi h})^{-\frac{d}{2}}\, \rho_{0,-h}~~~h\neq 0\nonumber\\
\rho_{gg} &\rightarrow&
(i)^{\frac{d}{2}}\rho_{g,-g}~~~g\neq 0\nonumber\\
\rho_{g,-g} &\rightarrow
& (-i)^{\frac{d}{2}}\rho_{-g,-g}~~~g\neq 0 \quad ,
\label{srho}
\eea
while under T-modular transformations ($\tau\rightarrow
\tau + 1$)
\bea
\Phi_{0,h} &\rightarrow&
(e^{-{i \pi/6}}) \, \Phi_{0,h}\nonumber\\
\Phi_{g,h} &\rightarrow&
(e^{i \pi/18}) \, \Phi_{g,g+h}~~~g\neq 0\nonumber\\
\label{tPhi}
\eea
and
\be
\eta^{-{D-2\over 2}} \rho_{gh}  \rightarrow
\eta^{-{D-2\over 2}} \rho_{g,g+h} \quad .
\label{tro}
\ee
The characters are combinations of chiral blocks that respect the
$Z_3$ group  structure.  In the $c=2$ case they are
\be
\chi_{\alpha,\beta}\, = \, {1 \over 3} \ \left[ \, \Phi_{\alpha,0} \, + \, 
\omega^{\beta} \Phi_{\alpha,1} \, + \, {\bar\omega}^{\beta} 
\Phi_{\alpha,2} \, \right] \qquad,
\label{charPhi}
\ee
while in the superstring models they are
\be
\chi_{\alpha,\beta}\, = \, {1 \over 3} \ \left[ \, \rho_{\alpha,0} \, + \, 
\omega^{\beta} \rho_{\alpha,1} \, + \, {\bar\omega}^{\beta} 
\rho_{\alpha,2} \, \right] \qquad,
\label{charro}
\ee
The modular transformation ${P}= ST^2ST$ relates
chiral traces in the transverse and direct M\"obius-strip amplitudes
and corresponds to $\widehat{P}= T^{1/2}ST^2ST^{1/2}$
on ``hatted'' real characters \cite{bs}, defined by 
\be
\hat\chi_{\alpha,\beta} \ = \ e^{- i \pi \, (h_{\alpha,\beta}-c/24)} 
\ \chi_{\alpha,\beta}
\label{hatted}
\ee
where $h_{\alpha,\beta}$ is the conformal weight of the character 
$\chi_{\alpha,\beta}$.

\rnc{\Large}{\normalsize}


\begin{thebibliography}{00}
\addcontentsline{toc}{section}{References}
\frenchspacing
\small
\addtolength{\itemsep}{-4pt}

\bibitem{cargese}
A.~Sagnotti,
ROM2F-87/25
{\it Talk presented at the Cargese Summer Institute on Non-Perturbative 
Methods in Field Theory, Cargese, Italy, Jul 16-30, 1987}.

\bibitem{ps}
G.~Pradisi and A.~Sagnotti,
Phys.\ Lett.\  {\bf B216}, 59 (1989).

\bibitem{bs}
M.~Bianchi and A.~Sagnotti,
Phys.\ Lett.\  {\bf B247}, 517 (1990);
Nucl.\ Phys.\  {\bf B361}, 519 (1991).

\bibitem{toroidal}
M.~Bianchi, G.~Pradisi and A.~Sagnotti,
Nucl.\ Phys.\  {\bf B376}, 365 (1992).

\bibitem{dien}
For a review see K.~R.~Dienes,
Phys.\ Rept.\  {\bf 287} (1997) 447
[hep-th/9602045].

\bibitem{mtheo}
For reviews see M.~J.~Duff,
Int.\ J.\ Mod.\ Phys.\  {\bf A11} (1996) 5623
[hep-th/9608117];
A.~Sen,
Nucl.\ Phys.\ Proc.\ Suppl.\  {\bf 58} (1997) 5
[hep-th/9609176].

\bibitem{pol}
J.~Polchinski,
Phys.\ Rev.\ Lett.\  {\bf 75} (1995) 4724
[hep-th/9510017].

\bibitem{tevscale}
For reviews see L.~E.~Ibanez,
hep-ph/9911499;
I.~Antoniadis,
hep-th/9909212;
I.~Antoniadis and A.~Sagnotti,
hep-th/9911205;
C.~Bachas,
hep-th/9907023.

\bibitem{kakutye}
Z.~Kakushadze and S.~H.~Tye,
Nucl.\ Phys.\  {\bf B548} (1999) 180
[hep-th/9809147].

\bibitem{sonoda}
H.~Sonoda,
Nucl.\ Phys.\  {\bf B311}, 401 (1988);
Nucl.\ Phys.\  {\bf B311}, 417 (1988).

\bibitem{lew}
D.~C.~Lewellen,
Nucl.\ Phys.\  {\bf B372}, 654 (1992).

\bibitem{fps}
D.~Fioravanti, G.~Pradisi and A.~Sagnotti,
Phys.\ Lett.\  {\bf B321}, 349 (1994)
[hep-th/9311183].

\bibitem{pss1}
G.~Pradisi, A.~Sagnotti and Y.~S.~Stanev,
Phys.\ Lett.\  {\bf B354}, 279 (1995)
[hep-th/9503207];
Phys.\ Lett.\  {\bf B356}, 230 (1995)
[hep-th/9506014].

\bibitem{pss2}
G.~Pradisi, A.~Sagnotti and Y.~S.~Stanev,
Phys.\ Lett.\  {\bf B381}, 97 (1996)
[hep-th/9603097].

\bibitem{susy95}
A.~Sagnotti,
hep-th/9509080.

\bibitem{fusc}
J.~Fuchs and C.~Schweigert,
Nucl.\ Phys.\  {\bf B530}, 99 (1998)
[hep-th/9712257];
Nucl.\ Phys.\  {\bf B558}, 419 (1999)
[hep-th/9902132]; 
hep-th/9908025.

\bibitem{gepner}
C.~Angelantonj, M.~Bianchi, G.~Pradisi, A.~Sagnotti and Y.~S.~Stanev,
Phys.\ Lett.\  {\bf B387}, 743 (1996)
[hep-th/9607229].

\bibitem{dapa}
A.~Dabholkar and J.~Park,
Nucl.\ Phys.\  {\bf B477}, 701 (1996)
[hep-th/9604178].

\bibitem{ads}
I.~Antoniadis, E.~Dudas and A.~Sagnotti,
Phys.\ Lett.\  {\bf B464}, 38 (1999)
[hep-th/9908023].

\bibitem{aadds}
C.~Angelantonj, I.~Antoniadis, G.~D'Appollonio, E.~Dudas and A.~Sagnotti,
hep-th/9911081.

\bibitem{resho}
A.~Recknagel and V.~Schomerus,
Nucl.\ Phys.\  {\bf B531}, 185 (1998)
[hep-th/9712186].

\bibitem{gijo}
E.~G.~Gimon and C.~V.~Johnson,
Nucl.\ Phys.\  {\bf B477}, 715 (1996)
[hep-th/9604129].

\bibitem{bgk}
R.~Blumenhagen, L.~Gorlich and B.~Kors,
hep-th/9908130.

\bibitem{ab}
C.~Angelantonj and R.~Blumenhagen,
hep-th/9911190.

\bibitem{noi}
M.~Bianchi, J.~F.~Morales and G.~Pradisi,
hep-th/9910228.

\bibitem{zurB}
Z.~Kakushadze, G.~Shiu and S.~H.~Tye,
Phys.\ Rev.\  {\bf D58}, 086001 (1998)
[hep-th/9803141].

\bibitem{carlo}
C.~Angelantonj,
hep-th/9908064;
hep-th/9909003.

\bibitem{hora}
P.~Horava,
Phys.\ Lett.\  {\bf B231} (1989) 251.

\bibitem{tdual}
For a review, see 
A.~Giveon, M.~Porrati and E.~Rabinovici,
Phys.\ Rept.\  {\bf 244}, 77 (1994)
[hep-th/9401139].

\bibitem{fixpo}
D.~Gepner,
Phys.\ Lett.\  {\bf B222}, 207 (1989);
J.~Fuchs, B.~Schellekens and C.~Schweigert,
Nucl.\ Phys.\  {\bf B461}, 371 (1996)
[hep-th/9509105];
Nucl.\ Phys.\  {\bf B473}, 323 (1996)
[hep-th/9601078].

\bibitem{gs}
M.~B.~Green and J.~H.~Schwarz,
Phys.\ Lett.\  {\bf B149}, 117 (1984);
Phys.\ Lett.\  {\bf B151}, 21 (1985).

\bibitem{aug}
A.~Sagnotti,
Phys.\ Lett.\  {\bf B294}, 196 (1992)
[hep-th/9210127].

\bibitem{chiral}
C.~Angelantonj, M.~Bianchi, G.~Pradisi, A.~Sagnotti and Y.~S.~Stanev,
Phys.\ Lett.\  {\bf B385}, 96 (1996)
[hep-th/9606169].

\bibitem{twoa}
S.~Ishihara, H.~Kataoka and H.~Sato,
Phys.\ Rev.\  {\bf D60}, 126005 (1999)
[hep-th/9908017].

\bibitem{kaku}
Z.~Kakushadze, G.~Shiu and S.~H.~Tye,
Nucl.\ Phys.\  {\bf B533}, 25 (1998)
[hep-th/9804092];
Z.~Kakushadze,
Phys.\ Lett.\  {\bf B455}, 120 (1999)
[hep-th/9904007];
Phys.\ Lett.\  {\bf B459}, 497 (1999)
[hep-th/9905033].

\bibitem{bnew}
R.~Blumenhagen, L.~Gorlich and B.~Kors,
JHEP {\bf 0001}, 040 (2000)
[hep-th/9912204].

\end{thebibliography}
\end{document}